\documentclass[aps,pre,twocolumn,showpacs,showkeys]{revtex4}

\usepackage{graphicx}
\usepackage{dcolumn}
\usepackage{bm}
\usepackage{amssymb}
\usepackage{amsmath}
\usepackage{array}
\usepackage{rotating}
\usepackage{color}
\usepackage{hyperref}

\begin{document}

\title{Stimulated coherent spontaneous emission in an FEL with `quiet' bunches}

\author{V.A.~Goryashko}
\email{vitaliy.goryashko@physics.uu.se}
 \altaffiliation[Also at ]
      {Institute for Radiophysics and Electronics of NAS of Ukraine}
\author{V. Ziemann}%
 \affiliation{%
Uppsala University}%

\date{April 17, 2012}

\begin{abstract}
For a planar FEL configuration we study stimulated coherent spontaneous emission driven by a gradient of the bunch current in the presence of different levels of noise in bunches. To perform a vast amount of simulations required for obtaining statistically valid results, we developed a memory and time efficient one-dimensional  simulation code based on the integral solution to a Klein-Gordon equation describing the  field evolution. The longitudinal granularity of the electron bunch density originating from shot noise is maintained throughout the analysis. Three-dimensional effects like transverse emittance and diffraction are taken into account in simulations via an effective FEL parameter calculated  from Xie's fitting formula. Calculations are performed for an FEL model with the SwissFEL injector bunch parameters. It turns out that a reduction of noise by several orders of magnitude below the level of shot noise is required to mitigate the noise effect. We propose a novel scheme that allows for formation of electron bunches with a reduced level of noise and  a high gradient of the current at the bunch tail to enhance coherent spontaneous emission. The presented scheme uses effects of noise reduction and controlled microbunching instability and consists of a laser heater, a shot noise suppression section as well as  a bunch compressor. The noise factor and microbunching gain with and without laser heater are estimated. We found that shot noise reduction by three orders of magnitude can be achieved for a finite transverse size electron bunch.
\end{abstract}



\maketitle

\section{Introduction}

Operating free electron lasers in the XUV and X-ray regions are usually based on the SASE (Self Amplified Spontaneous Emission) process starting from  spontaneous emission of undulator radiation initiated by shot noise~\cite{McNeil_2010}. The latter is a stochastic process and the SASE FEL produces a series of random superradiant spikes with  a large variation of intensities~\cite{SSY_2010}.  A method of direct FEL seeding by a coherent quantum laser operating in higher harmonic generation regime (HHG) as well as methods of high-gain and echo-enabled harmonic generation (HGHG and EEHG) in FELs have been proposed to overcome this deficiency~\cite{Thompson_2009,Yu_1990,Stupakov_2009}. Apart from these external seeding, an intrinsic seeding originating from coherent spontaneous emission (CSE) driven by the current gradient of electron bunches can be employed~\cite{Doria_1993,Jaroszynski_1993,Krinsky_1999,McNeil_1999}. However, shot noise competes with the coherent seeding and reduces the radiation pulses coherence and pulse to pulse reproducibility. To aid the FEL process recent papers~\cite{Gover_2009,Stupakov_2011} have proposed schemes to decrease the noise level below  the shot noise level resulting in the so-called `quiet' bunches. Specifically, a space-charge dominated interaction region followed by a dispersion section suppresses shot noise across a wide range of frequencies (the experiment~\cite{Musumeci_2011} demonstrated that after 1/4 plasma oscillation period the initial periodic THz modulation is completely washed out and one obtains a uniform temporal profile of the bunch density as predicted in~\cite{Gover_2009}). Recently, the shot noise suppression technique in the optical wave range was verified experimentally for the first time~\cite{Gover_experiment}.

In view of the shot noise reduction schemes, the intrinsic seeding
driven by the current gradient of `quiet' electron bunches becomes an attractive way of obtaining pulse to pulse stable stimulated CSE radiation without external seeding lasers. The temporal structure of the radiation pulses can be predetermined by preparing electron bunches with a special current distribution so that amplified radiation pulses will exhibit a reproducibility determined by that of the electron bunch current~\cite{McNeil_2002}.
For example, in the experiment~\cite{SUNSHINE} very stable from pulse to pulse radiation signals very observed from rectangular bunches with steep rise and fall of the particle distributions at the beginning and at the end of the bunch. Thus, our goal is to investigate the stimulated CSE in the presence of different levels of noise and determine the level of  reduction of shot noise required  to obtain radiation pulses with a well predetermined temporal structure.

In order to initiate stimulated CSE  electron bunch current gradients have to be sufficiently large, this implies that the particle distribution at the tail of bunches has to have a steep fall. To create bunches with such a distribution, we propose a novel bunch formation scheme that uses the effects of noise suppression and controlled microbunching instability~\cite{SSY_2002,Heifets_2002}. The latter is employed in a way similar to that used in the longitudinal space charge amplifier (LSCA)~\cite{SY_2010}. The principal difference from an LSCA will be that only the tail of the bunch is allowed to be sensitive to the microbunching instability whereas the main core of the bunch has to be stable against the instability. This can be realized by using a laser heater~\cite{SSY_2004,Huang_2004} with a partial overlap between electron and laser pulses such that the bunch tail remains unheated. Then, after a magnetic compressor the bunch will have the microbunched tail with a scale of the current gradient variation of the bunch core of the order of the variation of the laser used for heating. Further, a short noise suppression scheme has to be applied to produce `quiet' bunches  with high current gradients. Note that magnetic bunch compression used in~\cite{SUNSHINE} resulted in steep variations of the bunch particle density and allowed coherent spontaneous emission to dominate over incoherent one. However, a detailed study of the formation of electron bunches with special density distribution is beyond the scope of the present study and we will limit ourselves with order-of-magnitude estimates of the bunch properties as the bunch passes through the  laser heater, shot noise suppression section and bunch compressor.

CSE from `quiet' bunches is of interest not only for VUV or X-ray FELs but also  THz FELs may benefit from this approach. Moreover, a THz FEL with moderate relativistic bunches can be an excellent stand for testing shot noise suppression techniques and formation of bunches with high current gradients via controlled microbunching instability. So, we first develop a simple absolute stable method for solving the non-averaged equations of free-space and waveguide FELs and then apply our simulation technique to study lasing in an FEL seeded by `quiet`  bunches with high current gradients. We will perform simulations for a set of the electron bunch parameters obtained at the SwissFEL injector.

\section{Integral solution to the 1D Klein-Gordon equation}

Using a complete set of transverse modes the problem of
excitation of electromagnetic radiation by relativistic electron bunches moving in an undulator in the presence of a waveguide or medium can be reduced to a system of equations in the form of a 1D Klein-Gordon equation~\cite{Freund_book}. Therefore, we will build an integral solution to the following equation
\begin{equation}\label{Klein_Gordon_vector}
 \Bigl(\frac{\partial^2}{\partial z^2} - \frac{1}{c^2}\frac{\partial^2}{\partial t^2} -
                        k_{\perp}^2 \Bigr)\vec A_{\perp}   = -\frac{4\pi}{c} \vec I_{\perp},
\end{equation}
where $z$ and $t$ are the longitudinal coordinate and time, respectively. In 1D free-space approximation ($k_\perp \equiv 0$)  Eq.~\eqref{Klein_Gordon_vector} is the wave equation and $\vec A_\perp$ is nothing but the transverse component of the vector-potential, $\vec I_{\perp}$ is the current density averaged over the bunch cross-section. In general case, $k_\perp $ and $\vec A_{\perp}$ are the transverse wavenumber and mode amplitude resulting from an expansion of the vector-potential into a convenient  set of transverse modes (waveguide modes, Hermite-Gaussian modes, hybrid optical-waveguide modes); $\vec I_{\perp}$ is the effective current density dictated by a convolution of the physical current density with transverse modes.

We label the discrete pointlike electrons by the index $q = 1, \ldots , Q_e$, where $Q_e$ is the total number of electrons in the bunch. Note that $Q_e$ is different for different bunches because of shot noise. Choosing the axial coordinate $z$ to be the independent variable, we denote the arrival time of the $q$th electron at $z$ by $t_q(z)$, and the transverse and longitudinal velocities by $\vec v_{\perp|q}(z)$ and $v_{z|q}$, respectively. Then, the effective current density reads
\begin{equation}\label{eff_cuurent}
  \vec I_\perp(z,t) = \tilde{\varrho} \sum_{q}^{Q_e} \frac{\vec v_{\perp|q}(z)}{v_{z|q}(z)} \delta[t-t_q(z)],
\end{equation}
where $\delta(z)$ is the Dirac delta-function. In 1D free-space approximation the sources of radiation can be treated as charged disks of infinite transverse extent, with charge per unit area $\tilde \varrho$ ($\tilde \varrho < 0$).

In what follows, we omit the vector notations because in linearly or helically polarized undulators the transverse electromagnetic field can be described by a scalar equation of the Klein-Gordon type~\cite{Freund_book}.

In an FEL incoherent undulator radiation from relativistic electrons is about $\gamma^4$ times larger in the forward direction (with respect to the bunch propagation) than in the backward direction, where $\gamma$ is the average bunch energy in terms the rest mass. In the case of stimulated emission the forward radiation is also strongly dominant and we will take into account only it. Then, a solution to Eq.~\eqref{Klein_Gordon_vector} describing the field \textit{co-propagating} with bunches is
\begin{equation}\label{A_solution_formal}
  A_\perp(z,t) = \frac{1}{c}\int\limits_{-T_e}^{t} dt'
            \int\limits_{0}^{z} dz' G(z-z',t-t')I_\perp(z',t'),
\end{equation}
where the Green function satisfies the equation
\begin{equation*}
  \Bigl(\frac{\partial^2}{\partial z^2} - \frac{1}{c^2}\frac{\partial^2}{\partial t^2}- k_\perp^2\Bigl) G = -4\pi \delta(z-z')\delta(t-t')
\end{equation*}
and reads~\cite{Polyanin_2001}
\begin{multline}\label{Green_function}
   G(z-z',t-t') = 2\pi c\,U[c(t-t') - (z-z')] \times \\
   J_0 \bigl[k_\perp \sqrt{c^2(t-t')^2 - (z-z')^2} \bigr] .
\end{multline}
Here $T_e$ is the bunch duration (electrons enter the interaction region during the time interval from $-T_e$ to $0$), $J_0(t)$ is the Bessel function of the first kind, $U[t]$ is the unit-step function. Note that only radiation emitted at time $t-(z-z')/v_{\mathrm{ph}}$  by electrons in position $z'$ contributes to the radiation observed at position $z$ and time $t$ because of causality.  Here, $v_{\mathrm{ph}}$ is the phase velocity of the radiation field.

Using Eqs.~\eqref{eff_cuurent}-\eqref{Green_function} after some algebra we get the electric field of the form
\begin{multline}\label{E_field_general}
  E_\perp(z,t) \equiv
     - \frac{1}{c}\frac{\partial A_\perp}{\partial t} = \\
     -\frac{\tilde{\varrho}}{c^2} \sum_{q}^{Q_e}  \int\limits_{0}^{z} dz' \; \frac{v_{\perp|q}(z')}{v_{z|q}(z')} \,\frac{\partial G(z-z',t-t_q(z')]}{\partial t} \times \\
     U[t-t_q(z')]U[t_q(z') + T_e].
\end{multline}

From Eq.~\eqref{E_field_general} in 1D free-space approximation for given electron trajectories in a planar undulator we recover the well-known result~\cite{Krinsky_1999}
\begin{multline}\label{field_1D_final}
  E_\perp(z,t) =  -  \sum_{q}^{Q_e}\frac{4\pi \tilde{\varrho}\mathcal{K}
                 \gamma_q}{1+\mathcal{K}^2/2}  \sin[\omega_q(t-t_q^e-z/c)] \times \\
                 U[c(t-t_q^e) - z] U[z - \bar v_{z |q} (t-t_q^e)],
\end{multline}
where $\omega_q = 2 \gamma_{q}^2 k_u \bar v_{z |q}/(1+\mathcal{K}^2/2)$; $\mathcal{K} $ and $k_u$ are the undulator parameter and wavenumber, respectively; $t_q^e$, $\bar v_{z |q}$ and $\gamma_q$ are the entrance time, average longitudinal velocity and energy in units of the rest mass of $q$th electron, respectively.
We see that the electric field is a superposition of plane waves with different frequencies that depend on the electron energy. Equation~\eqref{field_1D_final} provides a simple insight into formation of coherent spontaneous emission in an FEL and is used later for verification of the numerical simulation scheme. In particular, for a monoenergetic electron pulse with a rectangular current profile Eq.~\eqref{field_1D_final} shows that a sequence of radiation pulses, which are one radiation cycle in duration, develops at each end of the electron pulse, see Fig.~1 and discussion in~\cite{McNeil_1999}.

A plane wave in~\eqref{field_1D_final} produced by $q$th electron at longitudinal position $z$ is nontrivial only if $t \in [t_q^e +z/c, t_q^e + z/{\bar v_{q|z}}]$. This result comes directly from the Green function~\eqref{Green_function} (recall that Eq.~\eqref{field_1D_final} is derived for $k_\perp = 0$) and it turns out that two unit-step functions in~\eqref{E_field_general} lead to redundant information. Keeping in mind the last observation one can check that these two unit-step functions in Eq.~\eqref{E_field_general} may be omitted without the loss of generality.

\section{Non-averaged FEL model: numerical algorithm}

Let us introduce the traditional set of dimensionless variables~\cite{Bonifacio_1989}
\begin{equation}\label{normalization}
  \zeta  = \frac{z}{\ell_g}, \quad
  \tau = \frac{z - \bar v_\| t}{\ell_g(1-\bar \beta_\|)},
  \quad
  F  = \frac{1}{4}\frac{|e| E_\perp}{m_e c^2} \,\frac{ \ell_g}{\rho}
        \frac{\mathcal{K} J\!J}{\gamma_{r}^2}.
\end{equation}
Gain length $\ell_g$ is related to dimensionless Pierce parameter $\rho$ as $\ell_g = 1/(2k_u \rho)$,  and $\rho$ reads
\begin{equation}
  \rho = \Bigl[\frac{1}{16} \frac{I_0}{I_\alpha}
    \frac{\mathcal{K}^2 J\!J^2}{\gamma_{r}^3 \sigma_r^2 k_u^2} \Bigr]^{1/3},
\end{equation}
where $\sigma_r$ is the rms transverse size of the electron beam. Here  $\gamma_{r}$ is the resonant energy in units of the rest mass $$
J\!J = J_0\Bigl(\frac{\mathcal{K}^2}{4+2\mathcal{K}^2}\Bigr) - J_1\Bigl(\frac{\mathcal{K}^2}{4+2\mathcal{K}^2}\Bigr)
$$
is the traditional $J\!J$-factor, $\bar v_\|$ is the average longitudinal velocity, $e$ and $m_e$ are the electron charge and mass, respectively.

Then, using the normalization~\eqref{normalization}, at the Compton limit $\rho \ll 1$ we get the dimensionless self-consistent system of equations in the form
\begin{equation}\label{self-consistent_real_field}
\begin{split}
 & \frac{\mathrm{d} \tau_q}{\mathrm{d}\zeta} = 2\rho \mu_q,
 \qquad
 \frac{\mathrm{d} \mu_q}{\mathrm{d}\zeta} =
                  - 4 F (\zeta,\tau_q) \sin[\zeta/(2\rho)],
 \\
 & F(\zeta, \tau) =  \frac{2}{\bar n_\|}  \sum_{q=1}^{Q_e}
                     \int_0^{\zeta}\limits \chi_q \sin\Bigl[\frac{\zeta}{2\rho}\Bigr]
  \, \frac{\partial G[\tau- \tau_q(\zeta'), \zeta -\zeta' ]}{\partial \tau} \mathrm{d}\zeta',
\end{split}
\end{equation}
where $\bar n_\|$ is the number of electrons per unit $\tau$ entering the interaction region at $\zeta = 0$ and $\chi_q$ is the charge weighting parameter $\chi_q = I_\perp(\zeta = 0, \tau)/I_{\perp, \mathrm{pk}}$, where $I_\perp(\zeta = 0, \tau)$ is the effective electron current with maximum value $I_{\perp, \mathrm{pk}}$. Note that deriving Eq.~\eqref{self-consistent_real_field} we ignored undulator harmonics since CSE at the third harmonic is about two orders of magnitude smaller than at the fundamental frequency (higher harmonics are even smaller) since the change in the bunch current over the harmonic period is smaller and hence the smaller the source term driving CSE~\cite{McNeil_2002}.

To avoid bulky expressions we will demonstrate implementation of the numerical algorithm for 1D free-space case ($k_\perp = 0$), then $\partial G[\tau- \tau_q(\zeta'), \zeta -\zeta' ]/\partial \tau = \delta[\tau- \tau_q(\zeta') - \zeta + \zeta']$. Generalization for $k_\perp \neq 0$ is straightforward. It is also advantageous to express in an explicit way that the electric field is co-propagating with electrons, to this end we define the complex amplitude by the relation  $F = \mathrm{Re} \{\tilde F \mathrm{e}^{i\omega_r(z/c-t)} \}$ , where $\omega_r = 2\omega_u \gamma_{r}^2/(1+\mathcal{K}^2)/2$ is the resonant frequency that corresponds to the case of ideal synchronism of bunches with the radiation field at the undulator entrance. Then, we can re-write Eq.~\eqref{self-consistent_real_field} as
\begin{equation}\label{self-consistent_McNeil}
\begin{split}
 & \frac{\mathrm{d} \tau_q}{\mathrm{d}\zeta} = 2\rho \mu_q,
 \quad
 \frac{\mathrm{d} \mu_q}{\mathrm{d}\zeta} =
                  - 2 \mathrm{Re} \{ \tilde F (\zeta,\tau_q) \mathrm{e}^{i\tau_q/(2\rho)}\bigl(1-\mathrm{e}^{i\zeta/\rho}\bigr),
 \\
 & \tilde F(\zeta, \tau) =  \frac{1}{\bar n_\|}  \sum_{q=1}^{Q_e}  \int_0^{\zeta}\limits e^{-i\frac{\tau+\zeta'-\zeta}{2\rho}}
  \, \delta[\tau + \zeta' -\zeta - \tau_q(\zeta')] \mathrm{d}\zeta'.
\end{split}
\end{equation}
Note that our system~\eqref{self-consistent_McNeil} is equivalent to the corresponding one of Krinsky~\cite{Krinsky_1999} and differs from that of McNeil et al.~\cite{McNeil_1999} by term $\bigl(1-\mathrm{e}^{i\zeta/\rho}\bigr)$, which is omitted in~\cite{McNeil_1999} and relevant in the case of short bunches~\cite{Krinsky_1999}.

For the numerical simulation we introduce a uniform mesh over $\zeta$ with fixed step $\Delta\zeta$ such that the $k$th node is located at $\zeta_k = k\,\Delta\zeta $. Then, we replace the integral from $0$ to $\zeta$ by a sum of integrals over small intervals $\Delta\zeta$
\begin{equation}
  \int_0^{\zeta} (\ldots) \mathrm{d}\zeta' \rightarrow
   \sum\limits_{k=0}^{K} \int\limits_{k\, \Delta\zeta}^{(k+1)\, \Delta\zeta} (\ldots) \mathrm{d}\zeta',
\end{equation}
and expand $\tau_q(\zeta')$ into a Taylor series around $k\, \Delta\zeta$ up to the linear term
\begin{multline}
  \tau_q(\zeta') \approx \tau_q(k\, \Delta\zeta) +
   \frac{\mathrm{d} \tau_q}{\mathrm{d}\zeta'}\Bigl |_{\zeta'= k\, \Delta\zeta} (\zeta ' - k\, \Delta\zeta)
   =
   \\
   \tau_q^{(k)} + 2\rho \mu_q^{(k)} (\zeta ' - k\, \Delta\zeta),
\end{multline}
where $\tau_q^{(k)} = \tau_q(k\, \Delta\zeta)$ and $\mu_q^{(k)} = \mu_q(k\, \Delta\zeta)$. Then, the field amplitude at $K$th node may be written as
\begin{multline}
  \tilde F(\zeta_K,\tau) =  \frac{1}{\bar n_\|}  \sum\limits_{k=0}^{K-1} \sum_{q=1}^{Q_e}
                 \!\! \!\! \int\limits_{k\, \Delta\zeta}^{(k+1)\, \Delta\zeta} \!\!\!\!\!\!
                 \exp\Bigr[-\frac{i (\tau+\zeta'- K\, \Delta\zeta)}{2\rho} \Bigr]\times
\\
        \, \delta\bigl[\tau + \zeta' - K\, \Delta\zeta - \tau_q^{(k)} - 2\rho \mu_q^{(k)}(\zeta ' - k\, \Delta\zeta)\bigr] \mathrm{d}\zeta'.
\end{multline}
Because of the Dirac Delta function, the integral over $\zeta'$ is nontrivial only if
\begin{equation}\label{nontrivial_fleid_condition}
  k\, \Delta\zeta \leq \zeta'(\tau) < (k+1)\, \Delta\zeta,
\end{equation}
where
\begin{equation}
  \zeta'(\tau) = \frac{\tau - K\, \Delta\zeta+ 2\rho \mu_q^{(k)} k\, \Delta\zeta - \tau_q^{(k)}}
           {2\rho \mu_q^{(k)}-1}.
\end{equation}
Note that $\zeta'$ depends on time $\tau$ and  we have to scan over $\tau$ to  calculate the integral as a function time. Hence, we introduce a mesh w.r.t $\tau$  with step $\Delta\tau$ and label nodes by subscript $j$ such that $\tau_j = j\, \Delta\tau$.
When inequalities~\eqref{nontrivial_fleid_condition} are fulfilled the  field amplitude in position $\zeta_k$ at time $\tau_j$ reads
\begin{multline}
  \tilde F(\zeta_K,\tau_j) = \frac{1}{\bar n_\|}  \sum\limits_{k=0}^{K-1} \times
                 \\ \sum_{q=1}^{Q_e}
                 \exp\Bigr[-\frac{i}{2\rho}  \frac{2\rho \mu_q^{(k)}[\tau_j - (K-k)\Delta\zeta] - \tau_q^{(k)}}{2\rho \mu_q^{(k)}-1} \Bigr].
\end{multline}
Note that $F(\zeta_{K+1},\tau_j)$ and $F(\zeta_K,\tau_j)$ are related by
\begin{multline}
  \tilde F(\zeta_{K+1},\tau_j) = \tilde F(\zeta_K,\tau_j-\Delta\zeta) +
  \\
                 \frac{1}{\bar n_\|} \sum_{q=1}^{Q_e}
                 \exp\Bigr[-\frac{i}{2\rho}  \frac{2\rho \mu_q^{(K)} \tau_j - \tau_q^{(K)}}{2\rho \mu_q^{(K)}-1} \Bigr].
\end{multline}
The last formula plays a critical role for fast numerical simulations without limitations on the computer memory  since it allows one to keep in the computer memory only $2Q_e$ positions of electrons and the field generated at the previous step over $\zeta$.

The equations of motion are integrated  using a combination of the predictor-corrector scheme and two-point Adams-Bashforth-Moulton method~\cite{Potter_1973}
\begin{equation}
\begin{split}
  & \mu_q^{K+1/2} \! = \! \mu_q^{K} - \Delta\zeta \,
                    \mathrm{Re}\{\tilde F(\zeta_K, \tau_q^{K})
                    \mathrm{e}\Bigl[\frac{i \tau_q^{K}}{2\rho} \Bigr] \bigl(1-\mathrm{e}^{i\zeta_K/\rho}\bigr)\}, \\
  & \tau_q^{K+1/2} = \tau_q^{K} + \Delta\zeta\, \rho \mu_q^{K},
  \\
  & \mu_q^{K+1} =
     \mu_q^{K} - \Delta\zeta \, \mathrm{Re}\Bigl\{ \exp\Bigl[\frac{i \tau_q^{K+1/2}}{2\rho} \Bigr] \bigl(1-\mathrm{e}^{i\zeta_{K+1/2}/\rho}\bigr)\times
     \\
  &  \hspace{1.5cm}\bigl[3\tilde F(\zeta_K, \tau_q^{K+1/2})
       - \tilde F(\zeta_{K-1}, \tau_q^{K+1/2})\bigr] \Bigr\},
  \\
  & \tau_q^{K+1} = \tau_q^{K} + 2\Delta\zeta\, \rho \mu_q^{K+1/2}.
\end{split}
\end{equation}

As we mentioned one needs to scan the condition~\eqref{nontrivial_fleid_condition} over time to get the field amplitude as a function of time. Then, the question is what the time interval of scanning should be. Electrons enter the interaction region from $-T_e$ to $0$, then assuming the constant velocity one finds that the radiation field is nonzero at position $z$ if $t_\mathrm{pulse}(z) \in  [z/c - T_e, z/\bar v_\|]$. In practice, this condition is exact since the tail of the bunch moves with constant average velocity $\bar v_\|$ (the tail experiences no radiation field because of the slippage) but the head starts to radiate field that propagates with the velocity of light. It is also clear that electron transit time $t_q$ is within interval $t_\mathrm{pulse}(z)$ and we need to calculate field amplitude $F$ only for interval $t_\mathrm{pulse}(z)$ at given $z$. The corresponding interval for dimensionless time is dictated by $\tau_\mathrm{min} = 0$ and $\tau_\mathrm{max} = \zeta + \tau_e$.
Thus, the developed technique has the advantage over finite-difference techniques~\cite{McNeil_1999,Maroli_2011,Fawley_2011} that the radiation field is calculated only over the time interval where it is nonzero, whereas the use of the finite-difference method for solving the excitation equation~\eqref{self-consistent_McNeil} would require to calculate the field in all points of a mesh over interval $\tau\in [0,\zeta_\mathrm{max} + \tau_e]$, thus increasing the calculation time and required computer memory. Moreover, if a full time-space structure of the field is not needed,  one can calculate the field only on a time mesh surrounding the bunch, thus even more decreasing the computational time.

\section{Stimulated  coherent spontaneous emission}

We will employ a model of the bunch with a rectangular shape of the  current profile and a uniform  random distribution of electrons. The rectangular current distribution may be used in simulations if the rise and fall of the bunch density occur at a scale much smaller than the FEL resonant wavelength that can be a case for the SwissFEL injector if the controlled microbunching is applied. To model shot noise we use the technique proposed in~\cite{McNeil_2003} that put the real electron ensemble into a correspondence to a smaller ensemble of macroparticles with the same statistical proprieties. In our simulations we employ a 1D non-averaged FEL model, however, the 3D effects are taken into account by using an effective value of the FEL parameter $\rho$ calculated with Xie's fitting formula~\cite{Xie_1995}. This formula accounts for the effects of emittance (`matched' bunch focusing is assumed) and energy spread as well as for diffraction. The later turns out to be the most severe factor of degradation of the FEL process such that the effective 3D FEL parameter, $\rho_{ef}$, is almost two times smaller than the 1D FEL parameter, $\rho$. Since the energy spread is accounted for by $\rho_{ef}$ only several thousands of  macroparticles are used in simulations. This fact along with our memory efficient mathematical algorithm results in fast simulations. Main SwissFEL injector and proposed FEL parameters are listed in Tables~I and II \cite{SwissFEL}.

\begin{tabular}{lcc}
\multicolumn{3}{c}{ }
 \\
 \multicolumn{3}{c}{Table I. SwissFEL injector parameters \label{MyTable1}}
 \\
 \multicolumn{3}{c}{\hphantom{111} after bunch compression. }
 \\
\hline\hline
           Parameter           &        Symbol        &  Value   \\
\hline
        Electron energy        & $\gamma_{r}m_ec^2$ & 250 MeV
\\
         Bunch charge          &                      & 200 pC
\\
      Bunch peak current       &        $I_0$         &  350 A
        \\
 Transverse rms   size         &      $\sigma_r$      & 55 $\mu$m
 \\
         Energy spread         &$\sigma_\gamma m_ec^2$& 64 keV
 \\
     Normalized emittance      & $\varepsilon_n  $    & 0.36 mm$\cdot\!$ mrad
 \\
       Bunch length        &     $T_b$            &    190 fsec
       \\
       \hline\hline
\end{tabular}

\begin{tabular}{lcc}
\multicolumn{3}{c}{ }
 \\
 \multicolumn{3}{c}{Table II. Main parameters of the FEL. \label{MyTable2}}
 \\
\hline\hline
           Parameter           &        Symbol        &  Value   \\
\hline
       Undulator period        &       $\lambda_u$     &   4 cm           \\
      Undulator parameter      &  $\mathcal{K}$        &     3.2           \\
   Number of undulator periods &       $N_u$          &      200          \\
 Normalized interaction length &   $\zeta_\mathrm{max}$     &     23.83      \\
        FEL wavelength         &   $\lambda_r$        & 0.511 $\mu$m
    \\
         FEL parameter         &   $\rho$             & 0.0095
\\
       Cooperation length      &    $L_c$             &   4.29 $\mu$m          \\
               Gain length     &    $\ell_g$          &      33.57 cm       \\
  Normalized bunch length      &      $\tau_b$        &        13.3           \\
 Effective number of electrons &  $\bar n_{\|}$       & $1.25\cdot\! 10^9 $\\
  Effective energy spread      &    $\mu_\gamma$      &     0.016
   \\
    Effective emittance        &  $\mu_\varepsilon$   &  0.0025
    \\
         Diffraction parameter &    $\mu_d$           &     2.6
 \\
  Effective FEL parameter      &   $\rho_{ef}$        & 0.0053
 \\
 \hline\hline
 \multicolumn{3}{c}{ }
\end{tabular}

\begin{figure}[t!]
  \centering %
  \includegraphics{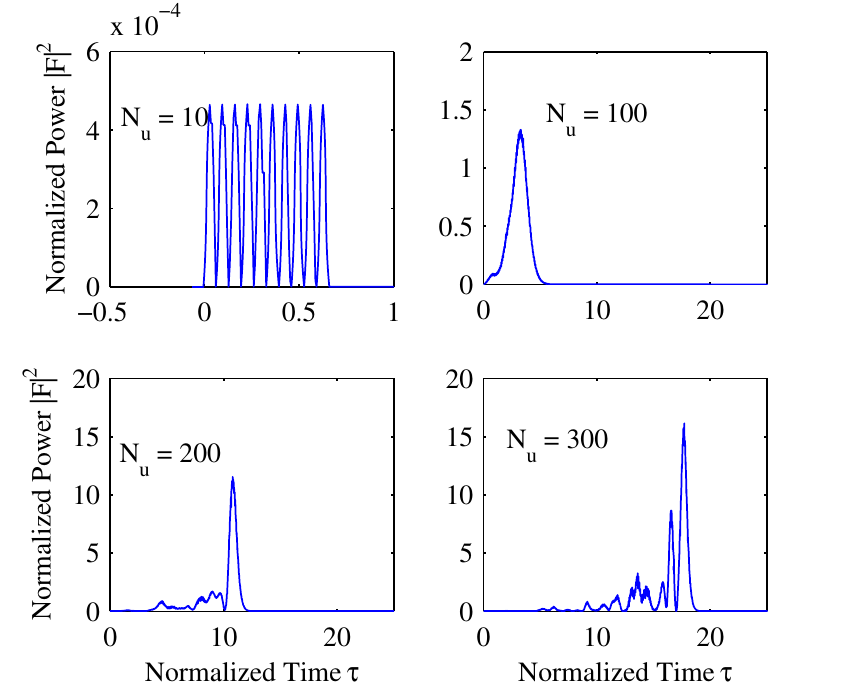}
\caption{\label{fig:Power_distribution_no_noise} Normalized power vs. normalized time at different cross-sections of the undulator. Shot noise is ignored in this simulation.}
\end{figure}
\begin{figure}[t!]
  \centering %
  \includegraphics{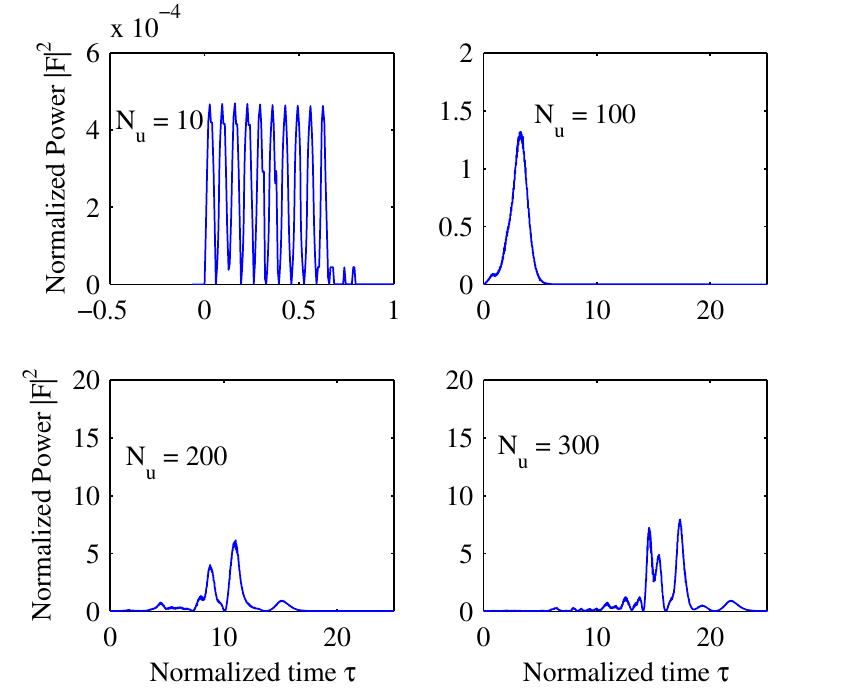}
\caption{\label{fig:Power_distribution_noise} Typical dependence of the normalized power on the normalized time at different cross-sections of the undulator with taking into account shot noise.}
\end{figure}

Coherent spontaneous radiation emitted by the bunch tail is amplified by electrons as it propagates through  the bunch resulting in the stimulated CSE, see Figs.~\ref{fig:Power_distribution_no_noise} and \ref{fig:Power_distribution_noise}. In Fig.~\ref{fig:Power_distribution_no_noise} we presented the radiation power as a function of time for an idealized case of the FEL without shot noise in the electron bunch. In the nonlinear regime the radiation pulse has a clear spike with a duration of the order of the cooperation length, $\Delta\tau \sim 1$ ($\Delta t \sim L_c/\bar v_\|$). This main spike leaves the bunch behind approximately in the cross-section  $\zeta \sim \tau_e  + 1$ ($z \sim \bar v_\| T_e + L_c$) and then it propagates without distortions in free-space.

Let us remark that in a THz FEL the cooperation length may achieve values smaller than $10\lambda_r$ such that the main spike of stimulated SCE  will be around ten cycle in duration or even shorter, therefore quite short (as compared to the resonant wavelength) pulses can be obtained by means of the FEL process itself.

Shot noise results in substantival distortions of the radiation pulse shape and reduction of the emitted power, see Fig.~\ref{fig:Power_distribution_noise}. To mitigate the shot noise effect several techniques have been proposed. However, to the authors knowledge the problem to what extent the bunch noise has to be suppressed to obtain well predetermined radiation pulses in the FEL was not studied and we address this question to our simulations of stimulated SCE.
\begin{figure}[b!]
  \includegraphics{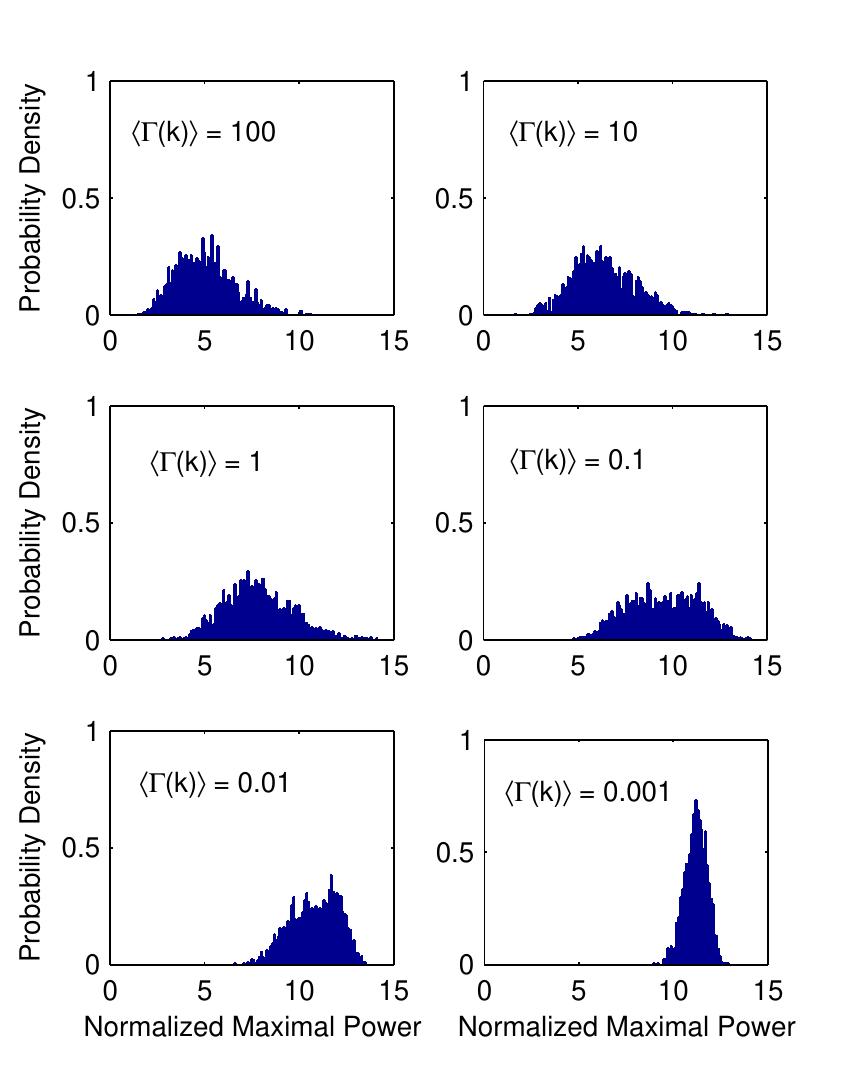}
\caption{\label{fig:probability_density} Histograms of the probability density distribution of the normalized maximal power $|F|^2$ for different levels of noise. Calculations have been performed with 1600 independent statistical events. The bunch and FEL parameters are given in Tables~I and II.}
\end{figure}

To characterize the level of noise at wavenumber $k$, we will follow the ref.~\cite{Stupakov_2011} and define the noise factor as
\begin{equation}
  \Gamma(k,s) = \frac{1}{Q_e} \sum\limits_{q,p} e^{i k [z_q(s) - z_p(s)]},
\end{equation}
where $z_q(s)$ is the longitudinal bunch coordinate of particle $q$ at position $s$ in the undulator. One can check that if the particle positions are uncorrelated, then shot noise results in $\langle\Gamma (k,s) \rangle = 1$, where $\langle \ldots \rangle$ stands for the statistical averaging. The situation $\langle\Gamma (k,s) \rangle < 1$ corresponds to the case of anticorrelated (`quiet') bunches. If the bunch energy spread is small, then magnetic compression causes the microbunching instability so that $\langle\Gamma (k,s) \rangle \sim G$, where $G$ is the microbunching gain. This situation corresponds to the correlated beams case.  In what follows, we will consider not only `quiet' bunches but also correlated ones since even in the presence of a laser heater the microbunching gain can greatly exceed unity at certain wavelength when a multistage compression is used.
\begin{figure}[b!]
  \includegraphics{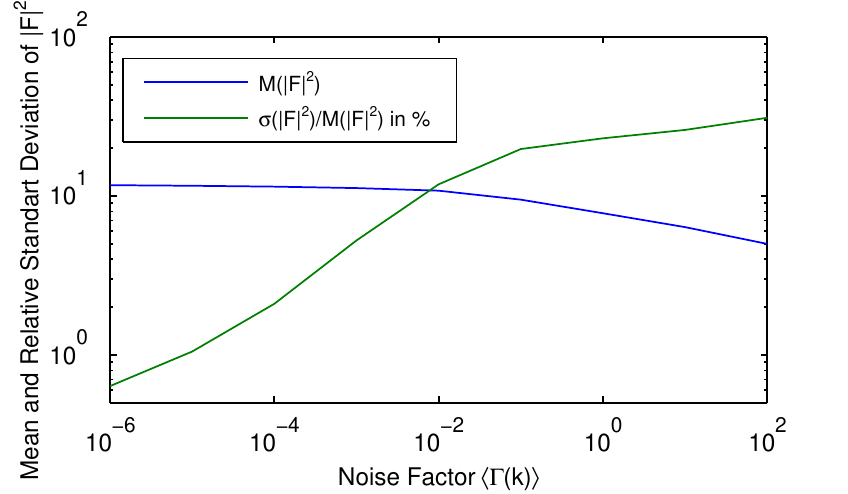}
\caption{\label{fig:mean} The mean and standard deviation of $|F|^2$ versus noise factor. Both axes are in a logarithmic scale.}
\end{figure}

\begin{figure}[t!]
  \includegraphics{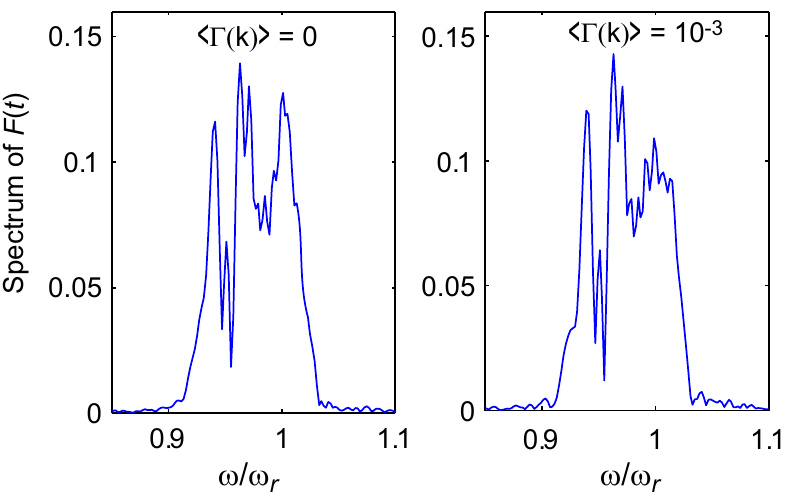}
\caption{\label{fig:fourier_field} Absolute value of the Fourier transform of $F(t)$ versus the frequency normalized to the resonant frequency, $\omega_r$.}
\end{figure}

To study the effect of noise on the FEL performance,  we calculated the density probability distribution of the maximal power of the pulse, see Fig.~\ref{fig:probability_density}. As one can expect, the mean maximal power, $\mathrm{M}(|F_{\mathrm{max}}|^2)$, increases and the relative standard deviation, $\sigma(|F_{\mathrm{max}}|^2)/\mathrm{M}(|F_{\mathrm{max}}|^2)$,  decreases as noise reduces, see Fig.~\ref{fig:mean}. The shape of the  density probability distribution also changes significantly. Noise with the level corresponding to shot noise almost halves the mean maximal power as compared to the case of zero noise level (7.8 of $|F_{\mathrm{max}}|^2$ against 11.7 of $|F_{\mathrm{max}}|^2$). In the latter case the maximal dimensional power is around 5.5~GW and the energy stored in the main radiation spike is around 100~$\mu$J. For $\langle\Gamma(k)\rangle = 10^{-3}$ the relative standard deviation of the maximal power from its value for $\langle\Gamma(k)\rangle = 0$ is less than 5\%. It turns out that for $\langle\Gamma(k)\rangle \ll 1$ the logarithm of the standard deviation, $\sigma(|F_{max}|^2)$, demonstrates quite linear dependence on $ \langle\Gamma(k)\rangle$ so that $\sigma(|F_{\mathrm{max}}|^2)$ is proportional to $ \langle\Gamma(k)\rangle^{1/3}$. We see that $\sigma(|F_{\mathrm{max}}|^2)$ diminishes slowly as $ \langle\Gamma(k)\rangle$ decreases and the reduction of the noise level by several orders of magnitude is required to obtain the relative standard deviation of the maximal power, let say, less than 10\%. A typical fourier transform of the normalized electric field for $ \langle\Gamma(k)\rangle = 0$ and  $ \langle\Gamma(k)\rangle = 10^{-3}$ is presented in Fig.~\ref{fig:fourier_field}. Analysis of numerous fourier transforms of $F(\tau)$ indicates that noise almost does not affect a lower frequency part of the spectrum but can distort a higher frequency part.

\section{Shot noise suppression and controlled microbunching instability}

In this section we propose and analyse a scheme that allows for formation of `quite' electron bunches with a high gradient of the current at the bunch tail. Let us consider the evolution of a `noisy' electron bunch from the injector exit to the undulator entrance allowing the bunch to pass through the  laser heater, shot noise suppression section and bunch compressor, see Fig.~\ref{fig:bunch_formation}. While the previous section describes the FEL performance in a rigorous way within 1D approximation, in this section we will give order-of-magnitude estimates of the relevant bunch characteristics during its formation in the proposed scheme since our intention to understand the physics of the process rather than to present technical specifications. More precise results for accurate technical specifications would require  start-to-end simulations from a photocathode to the FEL undulator entrance. For our analysis we will adopt typical approximations used for calculations of the microbunching gain and laser heater effect~\cite{Huang_2004}.

The temporal profile of the photocathode drive laser exhibits random fluctuations and they cause the density modulations of electron bunches. Because of this inhomogeneity of the electron density the space-charge oscillations are initiated and the conversion of density modulations into energy modulations as well as the reconversion occur each 1/4 plasma oscillation period. Then, at the injector exit electron bunches have some density modulation that can be characterized by a bunching factor
\begin{equation}
  b(k) = \frac{1}{Q_e e c}\int I(s) \mathrm{e}^{-i k s} ds,
\end{equation}
which is related to the noise factor defined, $\Gamma(k)$, by $\Gamma(k) = Q_e |b(k)|^2$. Here $I(s)$ is the bunch current. The microbunching gain is defined as a ratio of the bunching factors after and before compression. Following assumptions~\cite{Huang_2004}, in what follows we neglect the bunch energy modulation accumulated inside the injector and analyse the evolution of the electron density modulation starting from the injector end.

Let us estimate $\langle \Gamma(k)\rangle$ using Eqs.~(13), (22)-(24) from the ref.~\cite{Stupakov_2011} (note that the arguments of the Bessel functions should be $\sigma_r k/\gamma$ instead of the dimensional argument $k$).  The noise factor at the wavelength region from $\lambda_r/2$ to $3\lambda_r/2$ is presented in Fig.~\ref{fig:noise_suppression}. Recall that $\lambda_r$ is the FEL resonant wavelength. Because of the finite transverse size of electron bunches, $\langle \Gamma(k)\rangle$ depends on the wavelength and we chose dispersive strength $R_{56}$ to minimize $\langle \Gamma(k)\rangle$ at $\lambda = \lambda_r$. In our calculations, the drift section, $L_d$, is 10 meters. From Fig.~\ref{fig:noise_suppression} one can say that the noise reduction in the scheme employing the dispersive section~\cite{Stupakov_2011} is quite uniform as compared to the noise reduction obtained only by using the drift section of appropriate length~\cite{Gover_2011} (cf. our  Fig.~\ref{fig:noise_suppression} and Fig.~3 in \cite{Gover_2011}). In the later case, the noise reduction is observed only for a discrete set of wavelength. We conclude that the noise reduction by three orders of magnitude is achievable.
\begin{widetext}
\begin{center}
  \begin{figure}[b!]
  \includegraphics{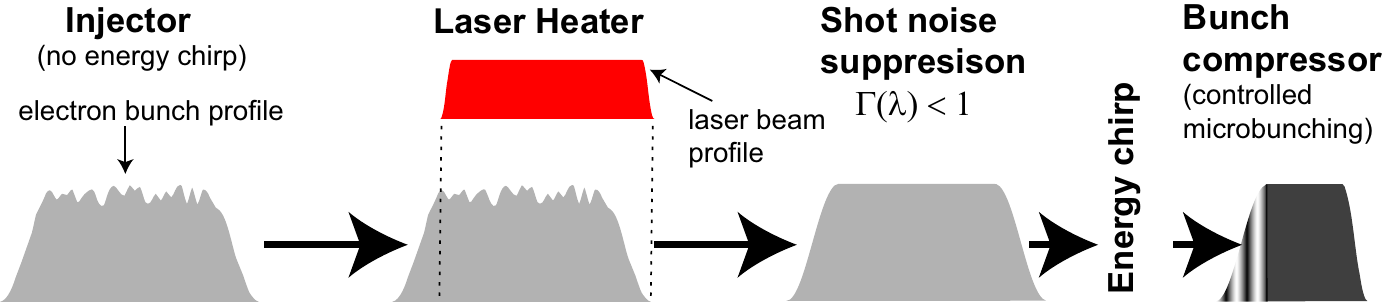}
\caption{\label{fig:bunch_formation} The schematic illustration of a possible formation of `quiet' bunch with a high current gradient at the bunch tail.}
\end{figure}
\end{center}
\end{widetext}

Because of impedances (mainly the space-charge impedance) the electron bunch density gives rise to the energy modulation  as bunches propagate though the drift section. This energy modulation is converted into an additional density modulation during magnetic compression, so-called microbunching gain. To avoid the microbunching gain, the laser heater that increases the uncorrelated energy spread thus providing strong Landau damping against the instability can be used. In Fig.~\ref{fig:microbunching_gain} we present the results of calculations of the microbunching gain for the system with and without laser heater. The maximal energy modulation, $\Delta\gamma_L m_e c^2$, produced by the laser heater has to be chosen in such a way that it results in an effective microbunching suppression on the one hand, and it does not affect the FEL performance because of the increased energy spread on the other hand. Without heating of electrons, an appreciable microbunching gain is observed approximately at a wavelength of 1 $\mu$m. Therefore, if we modulate the bunch tail with a period of 1 $\mu$m and heat with the laser heater only the main core of the bunch leaving the tail unheated, as it is shown in Fog.~\ref{fig:bunch_formation}, then we will get the microbunched tail and the steep variation of the electron density at the end of the main bunch core. Note that the microbunching gain in the vicinity of the resonant wavelength, which is $0.511\mu$m, is less than unity so that the reduced level of noise is preserved after magnetic compression.
\begin{figure}[t!]
  \includegraphics{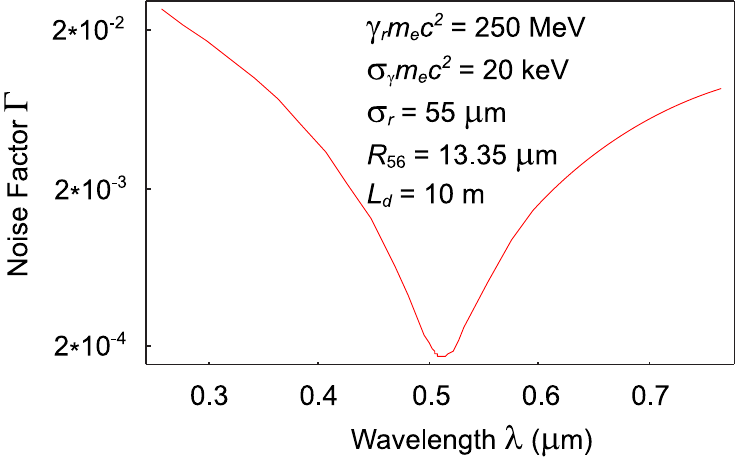}
\caption{\label{fig:noise_suppression} Noise factor versus wavelength for a bunch of the finite transverse size.}
\end{figure}
\begin{figure}[b!]
  \includegraphics{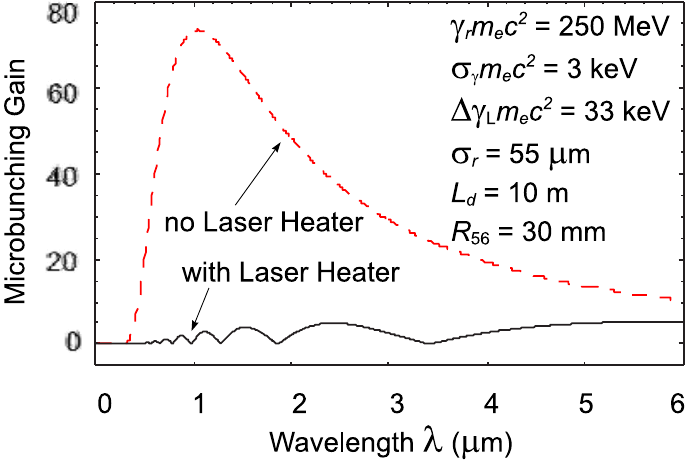}
\caption{\label{fig:microbunching_gain} Microbunching gain versus wavelength with and without the laser heater.}
\end{figure}

\section*{Conclusion}

We studied stimulated coherent spontaneous emission in a planar FEL driven by electron bunches with a rectangular shape of the charge density and different levels of density fluctuations. These density fluctuations originating from shot noise compete with an intrinsic coherent seeding driven by the current gradient and are in the focus of our investigation. We also proposed a scheme that allows for formation of electron bunches with a reduced level of noise and  a high gradient of the current at the bunch tail to enhance coherent spontaneous emission.

The need for a fast non-averaged time-dependent simulation code able to perform a vast amount of calculations required for obtaining statistically valid results stimulated us to develop such a code. The mathematical algorithm behind our code is conceptually similar to that behind the code `Fast'~\cite{SSY_book} and it employees an integral solution via the Green function to the 1D Klein-Gordon equation governing the electric field evolution. The algorithm presented is absolutely stable and does not impose constraints on discretization steps in longitudinal coordinate, observation time and entrance time of electrons. In the paper the field is calculated only on a mesh surrounding the bunch and then it is interpolated to the electron's positions, thus decreasing the computational time as compared to finite-difference methods that uses  a mesh depending on boundary conditions (advantages of integral methods over differential ones for injector design codes are discussed in details in~\cite{Giannessi_2003}). Our simulation code was verified against analytical result~\eqref{field_1D_final} in a low gain regime and against results presented in~\cite{McNeil_1999} for a deep nonlinear regime and excellent agreement is found. To account for 3D effects we used in simulations the effective FEL parameter calculated  from Xie's fitting formula. Therefore, our simulation code being 1D nevertheless takes into account transverse emittance and diffraction.

The structure of electromagnetic pulses in the studied FEL is mainly predetermined by the stimulated coherent spontaneous emission since the coherent part of spontaneous undulator radiation driven by the current gradient is dominant over the incoherent one associated with shot noise. However, the incoherent emission being small leads to essential distortions of radiation pulses during amplification and strongly affects the probability density distribution  of the maximal power, as shown in Fig.~\ref{fig:probability_density}.  It turns out that the standard deviation of $|F_{\mathrm{max}}|^2$ diminishes slowly with the reduction of noise because $\sigma(|F_{\mathrm{max}}|^2)$  is proportional to $\langle\Gamma(k)\rangle^{1/3}$ for $\langle\Gamma(k)\rangle \ll 1$. We found that shot noise should be reduced by at least three orders of magnitude to provide the relative standard deviation of the maximal output power less than 5\%. The FEL process starts from spontaneous undulator emission and the ratio of the power of  CSE to the power of incoherent emission is $Q_e |\mathcal{F}(\omega)|^2$, where $\mathcal{F}(\omega)$ is the form-factor of the bunch~\cite{Jaroszynski_1993}. Therefore, for FELs operating in other wavelength regions we may expect the standard deviation of the maximal output power to depend on the noise level in a similar way as we found, if the value $Q_e |\mathcal{F}(\omega)|^2$ is fixed. Thus, VUV or X-ray FELs based on stimulated CSE would require the reduction of shot noise by three or four orders of magnitude to have radiation pulses with a well predetermined time structure.

CSE is driven by the gradient of the bunch current and bunches with a steep rise of the electron density at the tail are preferable to drive the FEL. To create such bunches also having a reduced level of noise we proposed a scheme that uses effects of noise reduction and controlled microbunching instability, and consists of a laser heater, a shot noise suppression section as well as  a bunch compressor. In our scheme the tail of the bunch is allowed to be sensitive to the microbunching instability whereas the main core of the bunch has to be stable against the instability. This can be realized by using a laser heater with a partial overlap between electron and laser pulses such that the bunch tail remains unheated. Then, the bunch passes through the shot noise suppression scheme proposed in~\cite{Stupakov_2011} that consists of drift and dispersion sections, where the level of noise is reduced. We found that shot noise reduction by three orders of magnitude can be achieved for electron bunches produced by the SwissFEL injector. Then, in the bunch compressor the bunch undergoes the longitudinal compression and the  controlled microbunching of its tail occurs as well. We calculated the microbunching gain, $G$, with and without laser heater and in the latter case $G$ attains its maximal value around 80 at a wavelength of 1~$\mu$m (start-to-end calculations from the photocathode shows that $G$ can even achieve a value of several thousands~\cite{Simona}). Therefore, if we modulate the bunch tail with a period of 1 $\mu$m by means of a photocathode laser (or with an undulator and an external laser) and heat with the laser heater only the main core of the bunch leaving the tail unheated, as it is shown in Fog.~\ref{fig:bunch_formation}, then we will get the microbunched tail and the steep variation of the electron density at the end of the main bunch core. Note that the microbunching gain in the vicinity of the resonant wavelength, which is $0.511\mu$m, is less than unity so that the reduced level of noise is preserved after magnetic compression. Thus, one can form `quiet' bunches able effectively to drive CSE in the FEL.

\begin{acknowledgments}
The authors would like to acknowledge fruitful discussions with
Drs. M.~Dohlus, E.A.~Schneidmiller and S.~Bettoni. This publication has been produced during V.A.~Goryashko scholarship period at Uppsala University, thanks to the Swedish Institute for a  VISBY scholarship.
\end{acknowledgments}

\bibliography{goryashko}

\end{document}